\documentclass[%
 reprint,
superscriptaddress,
 amsmath,amssymb,
 aps,
prl,
]{revtex4-2}
\usepackage[dvipdfmx]{graphicx}% Include figure files
\usepackage{dcolumn}% Align table columns on decimal point
\usepackage{bm}% bold math
\usepackage{physics}
\usepackage{braket}
\usepackage{amsmath,amsfonts,mathtools,latexsym}
 \usepackage{amssymb}%白抜き文字(mathbb)
 \usepackage{amsbsy} % 太字のためのパッケージ
 \usepackage{xcolor}%xcolorパッケージを使用して色を変更する
 \usepackage{color}
 \newcommand{\me}{\mathrm{e}}
 \newcommand{\mi}{\mathrm{i}}
 \newcommand{\mS}{\mathrm{S}}

\begin{document}

% Use the \preprint command to place your local institutional report
% number in the upper righthand corner of the title page in preprint mode.
% Multiple \preprint commands are allowed.
% Use the 'preprintnumbers' class option to override journal defaults
% to display numbers if necessary
%\preprint{}

%Title of paper
\title{Scaling-Enhanced Rapid Readout of a Qubit Ensemble Assisted by\\
High-Frequency Detector Modes}

\author{Hiroki Nakabayashi}
\email{nakaba@iis.u-tokyo.ac.jp}
\affiliation{Department of Physics, The University of Tokyo, 5-1-5 Kashiwanoha, Kashiwa, Chiba 277-8574, Japan}
\affiliation{Analytical Quantum Complexity RIKEN Hakubi Research Team, RIKEN Center for Quantum Computing (RQC), 2-1 Hirosawa, Wako, Saitama 351-0198, Japan}
\author{Yuichiro Matsuzaki}
\email{ymatsuzaki872@g.chuo-u.ac.jp}
%\email[]{Your e-mail address}
%\homepage[]{Your web page}
%\thanks{}
%\altaffiliation{}
\affiliation{Department of Electrical, Electronic, and Communication Engineering, Chuo University, 1-13-27 Kasuga, Bunkyo-ku, Tokyo, 112-8551, Japan}
%Collaboration name if desired (requires use of superscriptaddress
%option in \documentclass). \noaffiliation is required (may also be
%used with the \author command).
%\collaboration can be followed by \email, \homepage, \thanks as well.
%\collaboration{}
%\noaffiliation

\date{\today}

\begin{abstract}
Quantum measurements of large qubit ensembles are often performed indirectly by coupling the ensemble to a detector and subsequently measuring the detector.
Despite the importance of rapid collective readout, it remains unclear what determines how the readout time scales with the number of qubits $N$.
Here, we first establish a benchmark $N^{-1/2}$ for a broad class of detectors without ultraviolet high-frequency modes.
We then show that the detector with unbounded high-frequency modes can surpass this benchmark and yield a characteristic time scaling $N^{-1/(2-\nu)}$ for \(0 < \nu < 2\), where $\nu$ characterizes the spectral structure of the detector.
Our results establish high-frequency detector modes as a resource for achieving a scaling advantage in collective quantum readout, with the detector spectrum directly controlling the scaling exponent of the readout time.
\end{abstract}

% insert suggested keywords - APS authors don't need to do this
%\keywords{}

%\maketitle must follow title, authors, abstract, and keywords
\maketitle

% body of paper here - Use proper section commands
% References should be done using the \cite, \ref, and \label commands
\paragraph{Introduction.---}
Quantum readout is a fundamental task in quantum science and technology \cite{helstrom1976quantum,holevo1982probabilistic,clerk2010introduction}.
It converts information stored in quantum degrees of freedom into an accessible classical record and thereby underlies quantum computation, quantum repeaters, and quantum thermodynamics \cite{elzerman2004single,pla2013high,sagawa2008second,cottet2017observing,masuyama2018information,briegel1998quantum}.
Ensemble measurements on qubits are particularly important in quantum sensing \cite{giovannetti2006quantum,giovannetti2011advances,wineland1992spin}, electron spin resonance (ESR) \cite{bienfait2016reaching}, nuclear magnetic resonance (NMR) \cite{bloch1946nuclear,hoult1976signal}, and quantum feedback control \cite{cox2016deterministic}.

In many architectures, the relevant information encoded in a qubit ensemble is not obtained via direct measurement of the qubit ensemble.
Instead, a collective observable of the ensemble is coupled to a probe, meter, or output field, which is subsequently measured \cite{hammerer2010quantum,zhang2012collective,chen2014cavity,kohler2017cavity,eisenach2021cavity}.
For many applications, the information must also be extracted rapidly \cite{fowler2012surface,heinsoo2018rapid,chen2023transmon}.
If the time required for the indirect readout, which we call readout time, is too long, it increases exposure to unwanted noise, delays feedback operations, and limits the repetition rate of the protocols.
Substantial effort has been devoted to improving detectors, amplifiers, and readout resonators on specific platforms for rapid readout \cite{walter2017rapid,gunyho2024single,sunada2024photon,jeffrey2014fast,spring2025fast}.

Despite extensive progress in quantum readout, a general principle governing how rapidly information can be transferred from a large qubit ensemble to a detector has not been established.
In particular, the scaling of the readout time with the number of qubits $N$ in the ensemble remains unclear.
In many studies, the system-size scaling provides a fundamental measure of performance and underlies quantum advantages in quantum metrology, quantum batteries and quantum heat engines \cite{huelga1997improvement,giovannetti2004quantum,matsuzaki2011magnetic,chin2012quantum,campaioli2017enhancing,campisi2016power,kamimura2022quantum}.
This raises the questions of whether a general scaling benchmark exists for collective readout and whether and how the scaling itself can be improved.

In this Letter, we address this question and show that unbounded high-frequency detector modes can provide a scaling advantage in collective readout.
We quantify the information transferred from the qubit ensemble to the detector by their quantum mutual information, which characterize the correlation between the qubit ensemble and the detector.
We first establish a lower bound on the readout time as $N^{-1/2}$ for detectors without ultraviolet high-frequency modes.
We then propose a new protocol that achieves a more favorable scaling by going beyond the $N^{-1/2}$ benchmark, using unbounded high-frequency detector modes.
In the idealized unbounded-frequency limit, the high-frequency modes make the coefficient appearing in the scaling law divergent, and the $N^{-1/2}$ benchmark no longer applies.
This leads to a different scaling law $N^{-1/(2-\nu)}$ for \(0 < \nu < 2\), where $\nu$ characterizes the high-frequency behavior of the detector spectral structure.
The resulting scaling is parametrically faster than the $N^{-1/2}$ scaling.

Concretely, we consider a spin-boson model in which a product state of $N$ qubits is collectively and longitudinally coupled to a bosonic detector \cite{leggett1987dynamics,weiss2012quantum,palma1996quantum}.
The qubit information is transferred to the detector through the collective spin-boson interaction.
We furthermore investigate a concrete readout protocol in which information about an unknown magnetic field is encoded in the qubit ensemble and the field is estimated by reading out the bosonic detector.
We show that the time required to attain the standard quantum limit (SQL), which is characterized by a estimation error of order $N^{-1/2}$, has the same $N$-dependence on time as the mutual-information timescale.

We also show that even when the detector has an ultraviolet cutoff, the scaling advantage persists over a finite range of system sizes.
The cutoff determines the range of system sizes over which the enhanced scaling can be observed.
Thus, the detector spectrum governs both the attainable scaling of the readout time and the range of system sizes over which this scaling advantage can be realized.
Our results identify high-frequency detector modes as a resource for accelerating collective quantum readout.

\paragraph{General benchmark for collective readout.---}

We here derive a broadly applicable benchmark for the time-scaling required to transfer information from the qubit ensemble to a detector without high-frequency modes.
In an indirect measurement, the interaction between the qubit ensemble and the detector generates correlations through which information about the ensemble is transferred to the detector.
We quantify this information transfer by the quantum mutual information between the qubit ensemble and the detector, and derive a lower bound on the time-scaling required to generate a fixed amount of such correlation.

For a regular detector, we first obtain the expected timescale $t\sim N^{-1/2}$ using a simple argument and then prove that it gives a rigorous lower bound on the readout time.
Throughout this paper, we use units with $\hbar=1$.
Suppose that the qubits couple to a detector through an additive observable of the qubit ensemble,
\begin{gather}
  H=H_{\rm S}\otimes I_{\rm D} +gO_{\rm S}\otimes B+I_{\rm S}\otimes H_{\rm D}, \label{eq:general-H}
\end{gather}
where $O_{\rm S}=\sum_{j=1}^N o_j$, $[H_{\rm S},O_{\rm S}]=0$, $H$ is the total Hamiltonian, $H_{\rm S}$ is the free Hamiltonian of the qubit ensemble, $g$ is a real constant, $B$ is a Hermitian operator acting on the detector, $H_{\rm D}$ is the free Hamiltonian of the detector, $o_j$ is the $N$-independent operator acting on the $j$-th qubit, and $I_{\rm S}$ and $I_{\rm D}$ are the identity operators on the qubit ensemble and detector, respectively.
The condition $[H_{\rm S},O_{\rm S}]=0$ implies $[O_{\rm S}\otimes I_{\rm D},H]=0$, so that $O_{\rm S}$ satisfies the quantum-nondemolition (QND) condition.
The spectral decomposition of $O_{\rm S}$ is given by $O_{\rm S}=\sum_m m P_m$, where $P_m$ is the projection operator onto the eigenspace of $O_{\rm S}$ with eigenvalue $m$.
We prepare an initial state as $\ket{\psi(0)}=\ket{s_0}\otimes \ket{d_0}$.
The state at time $t$ is given by
\begin{align}
  \ket{\psi(t)} &= \sum_{m: p_m>0} \sqrt{p_m}\,\ket{s_m(t)}\otimes\ket{d_m(t)}, \label{eq:general-state}
\end{align}
where $p_m:=\braket{s_0|P_m|s_0}$, $\ket{s_m}:=P_m \ket{s_0}/\sqrt{p_m}$, $\ket{s_m(t)}:=\me^{-\mi H_{\rm S}t}\ket{s_m}$, and $\ket{d_m(t)}:=\me^{-\mi(H_{\rm D}+gmB)t}\ket{d_0}$.

We first give a simple argument for the timescale for generating correlations between the qubit ensemble and the detector \cite{yang2018entanglement}.
For an initially pure product state, the short-time entanglement generation induced by the product interaction $gO_{\rm S}\otimes B$ is governed by the fluctuations of the two coupling operators: the characteristic scale governing the initial generation of entanglement \cite{yang2018entanglement} is $|g|\Delta O_{\rm S}\Delta B$, where $\Delta^2 O_{\rm S}:= \braket{s_0|O_{\rm S}^2|s_0} -\braket{s_0|O_{\rm S}|s_0}^2$ and $\Delta^2 B:= \braket{d_0|B^2|d_0} -\braket{d_0|B|d_0}^2$.
This quantity characterizes the strength with which the interaction generates correlations from the initial product state and has the dimension of an inverse time.
We therefore set the characteristic timescale as its inverse given by $(|g|\Delta O_{\rm S}\Delta B)^{-1}$.
When we assume a pure product input state $\ket{s_0}=\bigotimes_{j=1}^N\ket{\psi_j}$ and assume that the average local fluctuation remains finite and nonzero, the variance of the additive observable scales as $\Delta^2 O_{\rm S}=\Theta(N)$.
Therefore, when $g$ and $\Delta^2B$ are finite and independent of $N$, the timescale of generating correlations is $\Theta(N^{-1/2})$.
Thus, the characteristic timescale is expected to scale as $t\sim N^{-1/2}$ for generating correlations between the qubit ensemble and the detector.

We then show that the same $N^{-1/2}$ scaling gives a rigorous lower bound on the readout time by quantifying the quantum mutual information, which is an measure of the total correlation between the qubit ensemble and the detector generated by their interaction.
Since the total state is pure, the quantum mutual information is twice the entanglement entropy as $I_{\rm S:D}(t) =2S(\rho_{\rm D}(t))$, where $S(\rho):=-\Tr[\rho\log\rho]$ is the von Neumann entropy, $\rho_{\rm S}(t):=\Tr_{\rm D}[\ketbra{\psi(t)}]$ is the qubit state, and $\rho_{\rm D}(t):=\Tr_{\rm S}[\ketbra{\psi(t)}]$ is the detector state.

Whenever the maximum variance of $B$ under the conditional detector state $\ket{d_m(s)}$ defined as $V_{\rm B}:=\sup_{m, s\geq 0}\operatorname{Var}_{\ket{d_m(s)}}(B)$ is finite and independent of $N$, the quantum mutual information is bounded from above as follows:
\begin{gather}
  I_{\rm S:D}(t) \leq 2(C+1) \sqrt{|g|\sqrt{V_{\rm B}}\Delta O_{\rm S}t},\notag \\
  t_{\rm corr} \geq \frac{I_0^2} {4(C+1)^2|g|\sqrt{V_{\rm B}}\Delta O_{\rm S}} =\Theta(N^{-1/2}), \label{eq:general-time-bound}
\end{gather}
where \(t_{\rm corr}:=\inf\{t>0:I_{\rm S:D}(t)\geq I_0\}\), \(I_0>0\) and \(C\) are independent of \(N\) and \(t\), respectively \cite{supplemental}.
The correlation time defined above $t_{\rm corr}$ quantifies the time needed to generate a constant amount of correlation between the qubit ensemble and the detector for the quantum mutual information.

Therefore, we obtain the readout-time bound \(t_{\rm corr}=\Omega(N^{-1/2})\) whenever \(V_{\rm B}\) remains finite.
The finiteness of \(V_{\rm B}\) bounds the fluctuations of the detector coupling operator along all conditional detector states.
High-frequency detector modes can violate this condition due to the divergent of $V_{\rm B}$, for which Eq.~(\ref{eq:general-time-bound}) no longer gives a nonzero lower bound on \(t_{\rm corr}\).
We next determine the resulting readout timescale explicitly for a bosonic detector with such unbounded high-frequency modes.

\paragraph{High-frequency enhanced collective readout.---}
The $N^{-1/2}$ lower bound derived above is applicable for a broad class, but its discussion is based on a finite coefficient $V_{\rm B}$.
However, if we exploit unbounded high-frequency detector modes, $V_{\rm B}$ may diverge and the bound derived in Eq.~(\ref{eq:general-time-bound}) no longer applies.
We now propose a protocol to achieve a scaling advantage in the readout time in this case.
Concretely, we consider the bosonic detector with a ultraviolet component \cite{leggett1987dynamics,weiss2012quantum,palma1996quantum} and evaluate the lower and upper bounds on the quantum mutual information.
We show that both the lower and upper bounds are functions of $Nt^{2-\nu}$ for \(0 < \nu < 2\), which leads to the characteristic time scaling $t_{\rm corr}=\Theta(N^{-1/(2-\nu)})$.

Let us specialize Eq.~(\ref{eq:general-H}) to a bosonic detector consisting of a set of modes as follows:
\begin{gather}
  H=H_{\mS}\otimes I_{\rm D}+O_{\rm S}\otimes\sum_k g_k(b_k^\dagger+b_k)+I_{\mS}\otimes\sum_k\omega_k b_k^\dagger b_k, \label{eq:protocol-H}
\end{gather}
which satisfies the QND condition $\qty[H_{\rm S},O_{\rm S}]=0$.
Here, $b_k$ annihilates an excitation in detector mode $k$ of angular frequency $\omega_k$, and the parameter $g_k$ is the real coupling strength to mode $k$.
As discussed in Eq.~(\ref{eq:general-H}), the spectral decomposition of $O_{\rm S}$ is given by $O_{\rm S}=\sum_m mP_m$.
We prepare the detector in the vacuum and the qubits in the initial pure state $ \ket{\phi(0)} =\ket{s_0}\otimes\ket{\rm vac}_{\rm D}. $

By using the above setting, we obtain the reduced state of the detector in the interaction picture as follows:
\begin{gather}
  \rho_{\rm D}^{\rm I}(t)=\sum_m p_m \ketbra{\beta_m(t)}_{\rm D}, \notag\\
  \ket{\beta_m(t)}_{\rm D}:=\exp\qty[ m \sum_k g_k \qty(\alpha_k(t) b^\dagger_k - \alpha_k^*(t) b_k)]\ket{\rm vac}_{\rm D}, \label{eq:detector-coherent-state}
\end{gather}
where $p_m=\braket{s_0|P_m|s_0}$ and $\alpha_k(t):=(1-\me^{\mi\omega_k t})/\omega_k$ \cite{supplemental}.
Note that since the picture transformation is a local unitary operation, the quantum mutual information is invariant under the picture transformation.
The conditional coherent state $\ket{\beta_m(t)}_{\rm D}$ satisfies the following inner product:
\begin{gather}
  \abs{\braket{\beta_m(t)|\beta_{m'}(t)}}^2 =\me^{-A(t)(m-m')^2},\label{overlap}\\
  A(t):=\sum_k\abs{g_k}^2 \frac{2-2\cos(\omega_k t)}{\omega_k^2}. \label{eq:mode-def}
\end{gather}

We then derive bounds on the quantum mutual information $I_{\rm S:D}(t)$ for the product input $\ket{s_0}=\bigotimes_{j=1}^N\ket{\psi_j}$ and the additive observable $O_{\rm S}=\sum_{j=1}^N o_j$, where $o_j$ is independent of $N$.
Using the second Rényi entropy for the lower bound and the positivity of the quantum relative entropy for the upper bound, we obtain the following bounds on the quantum mutual information derived in the End Matter:
\begin{align}
  &\log\qty[ 1+4A(t)\Delta^2O_{\rm S} ]+o(1) \notag\\
  &\leq I_{\rm S:D}(t)\leq 2\qty[ 1+A(t)\Delta^2O_{\rm S} ] \log\qty[ 1+A(t)\Delta^2O_{\rm S} ]\notag\\
  &~~~~~~~~~~~~~~~ -2A(t)\Delta^2O_{\rm S} \log\qty[ A(t)\Delta^2O_{\rm S} ]. \label{eq:main-entropy-bounds}
\end{align}
For \(A(t)\Delta^2O_{\rm S}\to0\), the upper bound vanishes, and finite amount of qubits--detector correlation cannot be generated.
On the other hand, for \(A(t)\Delta^2O_{\rm S}=X_0\) with a fixed value $X_0$, both the lower and upper bound remains finite and positive in the large-\(N\) limit.
Therefore, for the product input with \(\Delta^2O_{\rm S}=\Theta(N)\), the characteristic correlation-formation scale is \(A(t)\Delta^2O_{\rm S}=\Theta(1)\), or equivalently \(A(t)N=\Theta(1)\).

We then evaluate the behavior of $A(t)$.
Assume the dispersion relation $\omega(\vec{k})=\Omega |\vec{k}|^n$ with $\Omega>0$ in $D$ spatial dimensions and a constant coupling $g$.
Here, starting from the finite discrete-mode model, we take the continuum and infinite-cutoff limits to model an ideal detector with unbounded high-frequency modes as follows:
\begin{align}
  A(t) &=\abs{g}^2S_D\int_{0}^\infty \dd k\, k^{D-1}\frac{2-2\cos(\Omega k^nt)}{\Omega^2k^{2n}}, \label{eq:main-A-cont}
\end{align}
where $S_D:=2\pi^{D/2}/\Gamma(D/2)$ is the surface area of a unit sphere in $D$ dimensions with $\Gamma(\cdot)$ being the gamma function, and $S_Dk^{D-1}\dd k$ is the radial integration measure.
For $\nu:=D/n$ with $0<\nu<2$, changing the integration variable to $u:=\Omega k^n t$, we obtain the fractional behavior \cite{nakabayashi2025fractional} of $A(t)$ as follows:
\begin{align}
  A(t) &= \frac{C_\nu|g|^2}{\Omega^\nu} t^{2-\nu}, \label{eq:main-frac-A}
\end{align}
where $ C_\nu ={\pi S_D/(n\Gamma(3-\nu)\sin(\pi \nu/2))}$.
Substituting Eq.~(\ref{eq:main-frac-A}) into $A(t)\Delta^2O_{\rm S}=\Theta(1)$ and using $\Delta^2O_{\rm S}=\Theta(N)$, we find that the characteristic readout timescale is therefore $ t_{\rm corr}=\Theta\left(N^{-1/(2-\nu)}\right). $
Since $1/(2-\nu)>1/2$, this time is shorter than the conventional timescale $\Theta(N^{-1/2})$.
The scaling advantage originates from the behavior of \(A(t)\), rather than from a change in the qubit fluctuations.
For a regular detector with $V_{\rm B}<\infty$, \(A(t)\propto t^2\), which yields the \(N^{-1/2}\) timescale.
On the other hand, unbounded high-frequency modes give \(A(t)\propto t^{2-\nu}\), leading to \(N^{-1/(2-\nu)}\).
The high-frequency spectral structure changes the exponent of the system-size scaling itself.

\paragraph{Photon-number readout and SQL time as a concrete protocol.---}
The mutual-information analysis above establishes the scaling of the time required to generate correlations between the qubit ensemble and the detector.
We now give an operational example based on the magnetic field estimation.
First, information about an unknown magnetic field $\omega_0$ is encoded in the spin state through its interaction with the spin ensemble.
The ensemble is then coupled to a bosonic detector, after which the photon number of an appropriately chosen detector mode is measured to estimate the magnetic field.
We show that, in this protocol, the time required to attain SQL sensitivity within a fixed constant factor has the same $N$-dependence as the mutual-information criterion discussed above.

Let us take the spin ensemble state encoded with the unknown parameter $\theta:=\omega_0 \tau$, where $\tau$ is the interrogation time, and the vacuum state of the detector as the initial state:
\begin{gather}
  \ket{\varphi (0)} =\qty(\cos\theta\ket{0}_{\mS} +\sin\theta\ket{1}_{\mS})^{\otimes N}\otimes\ket{\rm vac}_{\rm D}. \label{eq:encoded-state}
\end{gather}
We set the additive observable \(O_{\rm S}=J_z:=\sum_{j=1}^N\sigma_z^{(j)}\) in Eq.~(\ref{eq:protocol-H}), with \(\sigma_z\ket{1}=\ket{1}\) and \(\sigma_z\ket{0}=-\ket{0}\).
The eigenvalue \(m=2\ell-N\) occurs with the binomial probability \(q_\ell\), where \(\ell\) is the number of qubits in \(\ket{1}\).

The reduced detector state at time $t$ in the interaction picture is
\begin{align}
  \rho_{\rm D}^{\rm I}(t) &=\sum_{\ell=0}^Nq_\ell \ketbra{\beta_{2\ell-N}(t)}_{\rm D},\notag\\
  q_\ell&:=\binom{N}{\ell}\cos^{2(N-\ell)}\theta \times\sin^{2\ell}\theta.\label{eq:protocol-env}
\end{align}
The coefficient $q_\ell$ is the binomial probability of obtaining $\ell$ qubits in $\ket{1}_{\mS}$.

We then introduce the collective detector mode
\begin{gather}
  B_0(t)=\sum_k\frac{\alpha_k^*(t)g_k}{\sqrt{A(t)}}b_k, \qquad [B_0(t),B_0^\dagger(t)]=1,\label{eq:protocol-mode}
\end{gather}
and consider the photon-number measurement of this mode $\mathcal{N}(t):=B_0^\dagger(t)B_0(t)$.
The expectation value of the photon number operator is $\braket{\mathcal{N}(t)} :=\Tr_{\rm D}\qty[\rho_{\rm D}^{\rm I}(t)\mathcal{N}(t)]$.

We quantify the error for estimating $\theta$ given by $\Delta^2\theta:=\Delta^2\mathcal{N}(t)/\abs{\partial_{\theta}\braket{\mathcal{N}(t)}}^2$, where $\Delta^2\mathcal{N}(t):=\braket{\mathcal{N}^2(t)}-\braket{\mathcal{N}(t)}^2$ is the variance of the photon number operator.
Using the exact photon-number moments and the corresponding finite-\(N\) expression, we obtain the following result for large \(N\) \cite{supplemental}:
\begin{gather}
  \Delta^2\theta \simeq \frac{1}{4N} + \frac{1} {16A(t)N^2\sin^2(2\theta)}. \label{eq:main-error}
\end{gather}
Note that we assume that $\sin(4\theta)\neq 0$ to avoid the divergence.
The first term in Eq.~(\ref{eq:main-error}) gives the SQL contribution \(1/(4N)\), whereas the finite readout time produces the second term proportional to \(1/[A(t)N^2]\).
The latter becomes of order \(1/N\) when \(A(t)N=\Theta(1)\).
Thus, the photon-number measurement reaches SQL precision on the same characteristic timescale obtained above from the quantum mutual information.
More precisely, defining \(t_{\rm SQL}:=\inf\{t>0:\Delta^2\theta(t)\leq c/N\}\) with a constant \(c>1/4\), Eq.~(\ref{eq:main-error}) requires \(A(t_{\rm SQL})=\Theta(N^{-1})\).

For a regular detector with the conventional quadratic response $A(t)\propto t^2$, this condition gives $ t_{\rm SQL}=\Theta(N^{-1/2}). $
By contrast, for the unbounded high-frequency detector considered here, $A(t)\propto t^{2-\nu}$ from Eq.~(\ref{eq:main-frac-A}), and hence $ t_{\rm SQL}=\Theta\left(N^{-1/(2-\nu)}\right). $
Thus, for the same readout protocol and the same SQL-precision criterion, the high-frequency detector changes the scaling of the required measurement time from $N^{-1/2}$ to $N^{-1/(2-\nu)}$.
\paragraph{Effect of a finite cutoff.---}
\begin{figure}[tbp]
  \centering
  \includegraphics[width=\linewidth]{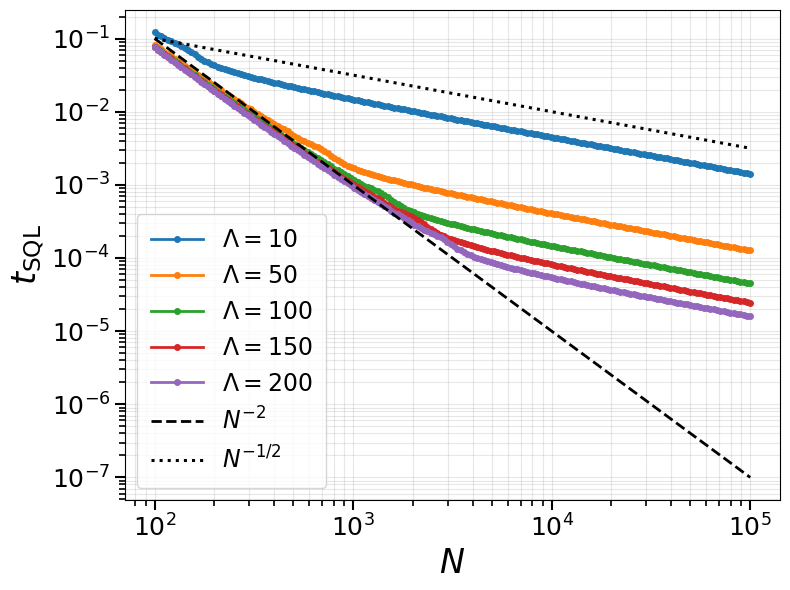}
  \caption{
Crossover of the readout time required for SQL precision, defined by $\Delta^2\theta(t_{\rm SQL})=1/N$, for $\theta=0.05$, $n=2$, $D=3$, $\Omega=1.0$, and $g=0.1$.
The solid curves show $t_{\rm SQL}$ for the finite cutoffs $\Lambda=10,~50,~100,~150,~200$, ordered from top to bottom, together with guide lines proportional to $N^{-2}$ and $N^{-1/2}$.}
  \label{fig:t-sql}
\end{figure}

The above analysis assumes an idealized infinite frequency range of the detector.
In practice, the detector has a finite bandwidth, which can be modeled by an ultraviolet wave-number cutoff $\Lambda$.
However, in the presence of a finite cutoff, the same scaling of the readout time is still observed for a finite range of $N$.
Let us introduce the ultraviolet cutoff $\Lambda$ in Eq.~(\ref{eq:main-A-cont}), i.e., truncate the integral at $\Lambda$.
Even with this cutoff, for times $t\gg (\Omega \Lambda^n)^{-1}$, the fractional behavior $A(t)\simeq C_\nu|g|^2t^{2-\nu}/\Omega^\nu$ in Eq.~(\ref{eq:main-frac-A}) is still valid \cite{supplemental}.
Therefore, the enhanced scaling $N^{-1/(2-\nu)}$ is observed if the enhanced scaling time is longer than the ultraviolet timescale, i.e., $t_{\rm corr},~t_{\rm SQL}\gg (\Omega \Lambda^n)^{-1}$.
Since the characteristic readout time scales as
\begin{gather}
  t_{\rm corr},~t_{\rm SQL}\sim \qty(\frac{\Omega^\nu}{C_\nu|g|^2N})^{1/(2-\nu)}
\end{gather}
up to a constant factor, this condition gives
\begin{gather}
  \Lambda\gg \Omega^{-1/n}\qty(\frac{C_\nu|g|^2N}{\Omega^\nu})^{1/[n(2-\nu)]}.
\end{gather}
Equivalently, for a fixed cutoff $\Lambda$, the enhanced scaling is observed for $N\ll N_{\rm c}$ with $N_{\rm c}\sim \Omega^2\Lambda^{n(2-\nu)}/(C_\nu|g|^2)$, see Fig.~\ref{fig:t-sql}.
On the other hand, for $N\gg N_{\rm c}$, since $A(t)$ behaves as a quadratic function of time in the relevant time regime, the scaling of the readout time returns to the conventional $N^{-1/2}$ scaling.

For a finite cutoff, the exponent \(\nu\) determines the enhanced system-size scaling, whereas the cutoff \(\Lambda\) determines the range of \(N\) over which this scaling can be observed.
Increasing \(\Lambda\) therefore shifts the crossover \(N_{\rm c}\) toward larger system sizes, while the asymptotic behavior at fixed finite \(\Lambda\) eventually returns to \(N^{-1/2}\).
In this sense, high-frequency modes control the range of system sizes over which the readout advantage is accessible.
Thus, the cutoff serves as a resource for rapid readout of a finite number of qubits.
\paragraph{Summary.---}
We first established a lower bound on the readout time of a qubit ensemble by quantifying the quantum mutual information between the qubit ensemble and the detector.
The lower bound is given by $t_{\rm corr}= \Omega(N^{-1/2})$ and is applicable to a broad class of detectors without ultraviolet high-frequency modes.
We then showed that high-frequency detector modes can accelerate collective readout to $t_{\rm corr}\propto N^{-1/(2-\nu)}$ with $0<\nu<2$.
These results establish high-frequency spectral structure as a resource for accelerating collective readout and suggest spectral engineering of detectors as a route toward fast measurement of large quantum ensembles.
The same accelerated scaling is attained by the photon-number readout and persists over a finite range of system sizes in the presence of a finite detector cutoff.

\begin{acknowledgments}
H.N. was supported by the RIKEN Junior Research Associate Program and the WINGS-QSTEP Program at the University of Tokyo.
%This project is supported by
Y.M. was supported by JST Moonshot R\&D Grant Number JPMJMS226C, JST CREST Grant Number JPMJCR23I5, and Presto JST Grant Number JPMJPR245B.
\end{acknowledgments}
% If you have acknowledgments, this puts in the proper section head.
%\begin{acknowledgments}
% put your acknowledgments here.
%\end{acknowledgments}

% Create the reference section using BibTeX:
\bibliography{rapidreadout}

\clearpage

\normalsize
\renewcommand{\theequation}{A\arabic{equation}}
\setcounter{equation}{0}

\onecolumngrid
\section*{End Matter}\label{sec:end_matter}
%%%%%%%%%%%%%%%%%%%%%%%%%%%%%%%%%%%%%%%%%%
\twocolumngrid
\paragraph{Lower bound on the quantum mutual information.---}
We evaluate the lower bound on the quantum mutual information in Eq.~(\ref{eq:main-entropy-bounds}) using the second R\'enyi entropy.
Since the second R\'enyi entropy is a lower bound on the von Neumann entropy \cite{renyi1961measures,renner2005security}, we have
\begin{align}
  I_{\rm S:D}(t) = 2S\qty(\rho_{\rm D}^{\rm I}(t)) \geq -2\log \Tr\qty[ \rho_{\rm D}^{\rm I}(t)^2 ]. \label{eq:EM-Renyi-lower-bound}
\end{align}
Using the coherent-state overlap in Eq.~(\ref{overlap}), the purity of the detector state is exactly given by
\begin{align}
  \Tr\qty[ \rho_{\rm D}^{\rm I}(t)^2 ] &= \sum_{m,m'} p_mp_{m'} \abs{\braket{\beta_m(t)|\beta_{m'}(t)}}^2 \notag\\
  &= \sum_{m,m'} p_mp_{m'} \exp\qty[ -A(t)(m-m')^2 ]. \label{eq:EM-detector-purity}
\end{align}

Taking $M$ to be a random variable distributed according to $p_m$, we rewrite Eq.~(\ref{eq:EM-detector-purity}) as
\begin{align}
  \Tr\qty[ \rho_{\rm D}^{\rm I}(t)^2 ] = \mathbb{E}\qty[ \exp\qty[ -A(t)(M-M')^2 ] ], \label{eq:EM-detector-purity-stochastic}
\end{align}
where $M'$ is an independent copy of $M$ and $\mathbb{E}[\cdot]$ denotes the expectation value.

We introduce the normalized random variable
\begin{gather}
  Z := \frac{ M-\braket{s_0|O_{\rm S}|s_0} }{ \sqrt{\Delta^2 O_{\rm S}} }.
\end{gather}
Equation~(\ref{eq:EM-detector-purity-stochastic}) then becomes
\begin{align}
  \Tr\qty[ \rho_{\rm D}^{\rm I}(t)^2 ] = \mathbb{E}\qty[ \exp\qty[ -A(t)\Delta^2O_{\rm S}(Z-Z')^2 ] ], \label{eq:EM-detector-purity-normalized}
\end{align}
where $Z'$ is an independent copy of $Z$.

For the product input $\ket{s_0}=\bigotimes_{j=1}^N\ket{\psi_j}$ and the additive observable $O_{\rm S}=\sum_{j=1}^N o_j$, the random variable $M$ is the sum of independent local measurement outcomes.
We here consider the case where the local operators are uniformly bounded and that
\begin{align}
  \Delta^2O_{\rm S} = \sum_{j=1}^N \qty[ \braket{\psi_j|o_j^2|\psi_j} -\braket{\psi_j|o_j|\psi_j}^2 ] =\Theta(N).
\end{align}
We consider an $N$-dependent time $t$ such that
\begin{gather}
  A(t)\Delta^2O_{\rm S}\to X_0
\end{gather}
as $N\to\infty$,
where $0<X_0<\infty$ is independent of $N$.
We then apply the central limit theorem to the normalized random variable $Z$, and the variables $Z$ and $Z'$ converge independently to standard normal random variables.
Therefore, in the large-$N$ limit, we have
\begin{align}
  &\lim_{N\to\infty} \Tr\qty[ \rho_{\rm D}^{\rm I}(t)^2 ]\notag\\
  &= \frac{1}{2\pi} \int_{-\infty}^{\infty}\dd z \int_{-\infty}^{\infty}\dd z' \exp\qty[ -X_0(z-z')^2 -\frac{z^2+z'^2}{2} ] \notag\\
  &= \frac{1}{\sqrt{1+4X_0}}.
\end{align}
Thus, the asymptotic behavior of the purity of the detector state is given by
\begin{align}
  \Tr\qty[ \rho_{\rm D}^{\rm I}(t)^2 ] = \frac{1}{\sqrt{1+4A(t)\Delta^2O_{\rm S}}}+o(1).
\end{align}

Therefore, the lower bound on the quantum mutual information in Eq.~(\ref{eq:EM-Renyi-lower-bound}) becomes
\begin{align}
  I_{\rm S:D}(t) &\geq -2\log\qty[ \frac{1}{\sqrt{1+4A(t)\Delta^2O_{\rm S}}}+o(1) ]\\
  &= \log\qty[ 1+4A(t)\Delta^2O_{\rm S} ]+o(1).
\end{align}

\paragraph{Upper bound on the quantum mutual information.---}
We next derive the upper bound in Eq.~(\ref{eq:main-entropy-bounds}).
When $A(t)=0$, all the conditional coherent states in Eq.~(\ref{eq:detector-coherent-state}) coincide with the detector vacuum, and the upper bound holds trivially.
For $A(t)>0$, we use the collective detector mode introduced in Eq.~(\ref{eq:protocol-mode}),
\begin{gather}
  B_0(t) =\sum_k\frac{\alpha_k^*(t)g_k}{\sqrt{A(t)}}b_k, \qquad [B_0(t),B_0^\dagger(t)]=1, \label{eq:EM-collective-detector-mode}
\end{gather}
where the commutation relation follows from the definition of $A(t)$ in Eq.~(\ref{eq:mode-def}).
In terms of $B_0(t)$, the conditional coherent state in Eq.~(\ref{eq:detector-coherent-state}) becomes
\begin{align}
  \ket{\beta_m(t)}_{\rm D} &=\exp\qty[ m\sqrt{A(t)} \left( B_0^\dagger(t)-B_0(t) \right) ]\ket{\rm vac}_{\rm D}. \label{eq:EM-conditional-collective-mode}
\end{align}
Thus, the dependence of the detector state on $m$ is entirely contained in the displacement of the collective detector mode $B_0(t)$.

Since the von Neumann entropy is invariant under unitary transformations, let us evaluate the detector state applying the displacement as follows:
\begin{align}
  &\ket{\widetilde{\beta}_m(t)}_{\rm D}\notag\\
  &:=\exp\qty[ -\braket{s_0|O_{\rm S}|s_0} \sqrt{A(t)} \left( B_0^\dagger(t)-B_0(t) \right) ] \ket{\beta_m(t)}_{\rm D} \\
  &= \exp\qty[ \left( m-\braket{s_0|O_{\rm S}|s_0} \right) \sqrt{A(t)} \left( B_0^\dagger(t)-B_0(t) \right) ]\notag\\
  &\quad \times \ket{\rm vac}_{\rm D}.
\end{align}
The displaced detector state is therefore
\begin{align}
  \widetilde{\rho}_{\rm D}^{\rm I}(t) &=\sum_m p_m \ketbra{\widetilde{\beta}_m(t)}_{\rm D}. \label{eq:EM-displaced-detector-state}
\end{align}
Using the invariance of the von Neumann entropy under unitary transformations, we have
\begin{align}
  S(\rho_{\rm D}^{\rm I}(t)) =S(\widetilde{\rho}_{\rm D}^{\rm I}(t)).
\end{align}

We now bound this entropy using the positivity of the quantum relative entropy.
For any $\lambda>0$, consider the thermal state of the collective mode $B_0(t)$, with all modes orthogonal to $B_0(t)$ in the vacuum as $ \sigma_\lambda = (1-\me^{-\lambda})\me^{ -\lambda B_0^\dagger(t)B_0(t) }\otimes \ketbra{\rm vac}_{\perp}$.

Using the positivity of the quantum relative entropy
\begin{align}
  D\qty( \widetilde{\rho}_{\rm D}^{\rm I}(t) \middle\| \sigma_\lambda ):=\Tr\qty[ \widetilde{\rho}_{\rm D}^{\rm I}(t) (\log\widetilde{\rho}_{\rm D}^{\rm I}(t)-\log\sigma_\lambda) ] \geq 0,
\end{align}
and on the support of $\widetilde{\rho}_{\rm D}^{\rm I}(t)$,
\begin{align}
  \log\sigma_\lambda = -\lambda B_0^\dagger(t)B_0(t) +\log(1-\me^{-\lambda}),
\end{align}
we obtain
\begin{align}
  S\qty(\widetilde{\rho}_{\rm D}^{\rm I}(t)) &=-D\qty( \widetilde{\rho}_{\rm D}^{\rm I}(t) \middle\| \sigma_\lambda )-\Tr\qty[ \widetilde{\rho}_{\rm D}^{\rm I}(t) \log\sigma_\lambda ] \notag\\
  &\leq \lambda \Tr\qty[ \widetilde{\rho}_{\rm D}^{\rm I}(t) B_0^\dagger(t)B_0(t) ] -\log(1-\me^{-\lambda}).
\end{align}
The first term is evaluated as
\begin{align}
  \Tr\qty[ \widetilde{\rho}_{\rm D}^{\rm I}(t) B_0^\dagger(t)B_0(t) ]&= \sum_m p_m \abs{ m-\braket{s_0|O_{\rm S}|s_0} }^2 A(t) \notag\\
  &= A(t)\Delta^2O_{\rm S}.
\end{align}
Therefore,
\begin{align}
  S\qty(\widetilde{\rho}_{\rm D}^{\rm I}(t)) &\leq \lambda A(t)\Delta^2O_{\rm S} -\log\qty(1-\me^{-\lambda}). \label{eq:EM-detector-entropy-lambda-bound}
\end{align}
Minimizing the right-hand side with respect to $\lambda$ for $A(t)\Delta^2O_{\rm S}>0$, we find the optimal value
\begin{align}
  \lambda = \log\qty[ \frac{ 1+A(t)\Delta^2O_{\rm S} }{ A(t)\Delta^2O_{\rm S} } ],
\end{align}
and hence
\begin{align}
  S\qty(\rho_{\rm D}^{\rm I}(t)) &= S\qty(\widetilde{\rho}_{\rm D}^{\rm I}(t)) \notag\\
  &\leq \qty[ 1+A(t)\Delta^2O_{\rm S} ] \log\qty[ 1+A(t)\Delta^2O_{\rm S} ] \notag\\
  &\quad - A(t)\Delta^2O_{\rm S} \log\qty[ A(t)\Delta^2O_{\rm S} ]. \label{eq:EM-detector-entropy-upper-bound}
\end{align}

Therefore, we obtain the upper bound on the quantum mutual information in Eq.~(\ref{eq:main-entropy-bounds}) as
\begin{align}
  I_{\rm S:D}(t) &\leq 2\qty[ 1+A(t)\Delta^2O_{\rm S} ] \log\qty[ 1+A(t)\Delta^2O_{\rm S} ]\notag\\
  &\quad -2A(t)\Delta^2O_{\rm S} \log\qty[ A(t)\Delta^2O_{\rm S} ].
\end{align}
\end{document}

\renewcommand{\theequation}{S\arabic{equation}}

\title{Supplemental Material for ``Scaling-Enhanced Rapid Readout of a Qubit Ensemble Assisted by High-Frequency Detector Modes''}

\author{Hiroki Nakabayashi}
\affiliation{Department of Physics, The University of Tokyo, 5-1-5 Kashiwanoha, Kashiwa, Chiba 277-8574, Japan}
\affiliation{Analytical Quantum Complexity RIKEN Hakubi Research Team, RIKEN Center for Quantum Computing (RQC), 2-1 Hirosawa, Wako, Saitama 351-0198, Japan}
\author{Yuichiro Matsuzaki}
\affiliation{Department of Electrical, Electronic, and Communication Engineering, Chuo University, 1-13-27 Kasuga, Bunkyo-ku, Tokyo, 112-8551, Japan}

\maketitle

\section{Derivation of the general benchmark (\ref{main-eq:general-time-bound}) for the quantum mutual information}

We begin with the Hamiltonian used in the main text in Eq.~(\ref{main-eq:general-H}):
\begin{gather}
  H=H_{\rm S}\otimes I_{\rm D} +gO_{\rm S}\otimes B+I_{\rm S}\otimes H_{\rm D}, \label{eq:general-H}
\end{gather}
where $O_{\rm S}=\sum_{j=1}^N o_j$, $[H_{\rm S},O_{\rm S}]=0$, $H$ is the total Hamiltonian, $H_{\rm S}$ is the free Hamiltonian of the qubit ensemble, $g$ is a real constant, $B$ is a Hermitian operator acting on the detector, $H_{\rm D}$ is the free Hamiltonian of the detector, $o_j$ is the $N$-independent operator acting on the $j$-th qubit, and $I_{\rm S}$ and $I_{\rm D}$ are the identity operators on the qubit ensemble and detector, respectively.
The condition $[H_{\rm S},O_{\rm S}]=0$ implies $[O_{\rm S}\otimes I_{\rm D},H]=0$, so that $O_{\rm S}$ satisfies the quantum-nondemolition (QND) condition.
The spectral decomposition of $O_{\rm S}$ is given by $O_{\rm S}=\sum_m m P_m$, where $P_m$ is the projection operator onto the eigenspace of $O_{\rm S}$ with eigenvalue $m$.

We prepare an initial state as $\ket{\psi(0)}=\ket{s_0}\otimes \ket{d_0}$.
The state at time $t$ is given by
\begin{align}
  \ket{\psi(t)} &=\sum_m \me^{-\mi H_{\rm S}t}P_m \ket{s_0} \otimes\me^{-\mi(H_{\rm D}+gmB)t}\ket{d_0}\\
  &:= \sum_{m: p_m>0} \sqrt{p_m}\,\ket{s_m(t)}\otimes\ket{d_m(t)}, \label{eq:general-state}
\end{align}
where $p_m:=\braket{s_0|P_m|s_0}$, $\ket{s_m}:=P_m \ket{s_0}/\sqrt{p_m}$, $\ket{s_m(t)}:=\me^{-\mi H_{\rm S}t}\ket{s_m}$, and $\ket{d_m(t)}:=\me^{-\mi(H_{\rm D}+gmB)t}\ket{d_0}$.

The states $\ket{s_m(t)}$ for different $m$ are orthonormal.
Consequently, the reduced detector state is
\begin{align}
  \rho_{\rm D}(t) =\sum_m p_m\ketbra{d_m(t)}{d_m(t)}. \label{eq:benchmark-detector-state}
\end{align}
Since the total state in Eq.~(\ref{eq:general-state}) is pure, its quantum mutual information is twice the von Neumann entropy of the reduced state:
\begin{align}
  I_{\rm S:D}(t) =S(\rho_{\rm S}(t))+S(\rho_{\rm D}(t))-S(\ketbra{\psi(t)}) =2S(\rho_{\rm S}(t))=2S(\rho_{\rm D}(t)), \qquad S(\rho):=-\Tr(\rho\log\rho). \label{eq:benchmark-pure-qmi}
\end{align}

We next derive an upper bound on $S(\rho_{\rm D}(t))$ that does not depend on the number of eigenvalues of $O_{\rm S}$.
For any positive operator $X$ whose support contains that of $\rho_{\rm D}(t)$, the positivity of the quantum relative entropy $D(\cdot\|\cdot)$ gives as follows:
\begin{align}
  D\left(\rho_{\rm D}(t)\middle\|\frac{X}{\Tr X}\right) := -S(\rho_{\rm D}(t)) -\Tr[\rho_{\rm D}(t)\log X] +\log\Tr X\geq 0.
\end{align}
Therefore,
\begin{align}
  S(\rho_{\rm D}(t)) \leq \log\Tr X -\Tr[\rho_{\rm D}(t)\log X]. \label{eq:benchmark-relative-entropy}
\end{align}
We now choose $X$ so that the second term can be bounded by the fluctuation $\Delta O_{\rm S}$ while keeping $\Tr X$ finite independently of the number of eigenvalues.
Let us take $X$ as
\begin{gather}
  X=\frac{1}{2}\sum_{j=0}^{K}\abs{A_{j+1}-A_j},
\end{gather}
where
\begin{gather}
  A_j=f_jQ_j, \qquad f_j:=\exp\left(-\frac{\abs{\mu_j-\braket{s_0|O_{\rm S}|s_0}}}{a}\right), \qquad Q_j:=\ketbra{d_{\mu_j}(t)}{d_{\mu_j}(t)}
\end{gather}
with a positive constant $a>0$.
Here $\mu_1<\mu_2<\cdots<\mu_K$ denote the eigenvalues of $O_{\rm S}$ with $p_{\mu_j}>0$, and we take $A_0=A_{K+1}:=0$.

We will derive the following relation by taking $X$ as above:
\begin{align}
  S(\rho_{\rm D}(t)) \leq \log\Tr X -\sum_{k=1}^{K}p_{\mu_k}\log f_k. \label{eq:benchmark-entropy-X}
\end{align}
Using the following relations
\begin{align}
  A_k = \sum_{j=0}^{k-1}(A_{j+1}-A_j),\quad A_k = -\sum_{j=k}^{K}(A_{j+1}-A_j),
\end{align}
we obtain
\begin{align}
  \sum_{j=0}^{k-1}\abs{A_{j+1}-A_j} \geq A_k,\quad \sum_{j=k}^{K}\abs{A_{j+1}-A_j} \geq A_k.\label{eq:benchmark-X-majorization}
\end{align}
Therefore, adding Eqs.~(\ref{eq:benchmark-X-majorization}) gives
\begin{gather}
  X=\frac{1}{2}\sum_{j=0}^{K}\abs{A_{j+1}-A_j}\geq A_k=f_kQ_k,
\end{gather}
for all $k=1,\ldots,K$.
Therefore, we obtain the inequality (\ref{eq:benchmark-entropy-X}) as follows:
\begin{align}
  S(\rho_{\rm D}(t)) &\leq \log\Tr X -\Tr[\rho_{\rm D}(t)\log X]\\
  &\leq \log\Tr X-\sum_{k=1}^{K}p_{\mu_k}\bra{d_{\mu_k}(t)}\log X\ket{d_{\mu_k}(t)}\\
  &\leq \log\Tr X-\sum_{k=1}^{K}p_{\mu_k}\log f_k
\end{align}

We then evaluate $\Tr X$ in Eq.~(\ref{eq:benchmark-entropy-X}).
By definition of $X$ and the triangle inequality, we have
\begin{align}
  \Tr X &= \frac12 \left[ f_1+f_K +\sum_{j=1}^{K-1} \left\| f_{j+1}Q_{j+1}-f_jQ_j \right\|_1 \right]\\
  &\leq \frac{1}{2}\qty[f_1+f_K+\sum_{j=1}^{K-1}\sqrt{(f_j-f_{j+1})^2+4f_jf_{j+1}\qty(1-\abs{\braket{d_{\mu_j}(t)|d_{\mu_{j+1}}(t)}}^2)}]. \label{eq:benchmark-trace-X}
\end{align}

Whenever the maximum variance of $B$ defined as
\begin{gather}
  V_{\rm B} := \sup_{m,\,s\geq0} \left[ \braket{d_m(s)|B^2|d_m(s)} -\braket{d_m(s)|B|d_m(s)}^2 \right] <\infty. \label{eq:benchmark-VB}
\end{gather}
is finite, the infidelity between the conditional detector states $\ket{d_m(t)}$ and $\ket{d_{m'}(t)}$ is bounded from above as follows:
\begin{align}
  1-\abs{\braket{d_m(t)|d_{m'}(t)}}^2 &\leq g^2V_{\rm B}(m-m')^2t^2.\label{eq:benchmark-overlap-bound}
\end{align}
This inequality can be checked from the time derivative of the overlap.
Differentiating the overlap, we have
\begin{align}
  \frac{\dd}{\dd t} \abs{\braket{d_m(t)|d_{m'}(t)}}^2 &\geq -2\abs{g(m-m')} \abs{\braket{d_m(t)|d_{m'}(t)}}\times \abs{ \braket{ d_m(t)| \left( B-\braket{d_{m'}(t)|B|d_{m'}(t)} \right) |d_{m'}(t)} }. \label{eq:benchmark-overlap-derivative}
\end{align}
Since
\begin{align}
  \braket{ d_{m'}(t)| \left( B-\braket{d_{m'}(t)|B|d_{m'}(t)} \right) |d_{m'}(t)} =0,
\end{align}%
the last factor in Eq.~(\ref{eq:benchmark-overlap-derivative}) is bounded as
\begin{align}
  \abs{ \braket{ d_m(t)| \left( B-\braket{d_{m'}(t)|B|d_{m'}(t)} \right) |d_{m'}(t)} } &=\abs{ \braket{ d_m(t)|\qty(1-\ketbra{d_{m'}(t)}{d_{m'}(t)}) \left( B-\braket{d_{m'}(t)|B|d_{m'}(t)} \right) |d_{m'}(t)} }\\
  &\leq \norm{ \qty(1-\ketbra{d_{m'}(t)}{d_{m'}(t)}) \ket{d_m(t)} } \norm{ \qty( B-\braket{d_{m'}(t)|B|d_{m'}(t)} ) \ket{d_{m'}(t)} }\\
  &= \sqrt{ 1-\abs{\braket{d_m(t)|d_{m'}(t)}}^2 }\, \sqrt{ \braket{ d_{m'}(t)| \left( B-\braket{d_{m'}(t)|B|d_{m'}(t)} \right)^2 |d_{m'}(t)} }\\
  &\leq \sqrt{ 1-\abs{\braket{d_m(t)|d_{m'}(t)}}^2 }\sqrt{V_{\rm B}}. \label{eq:benchmark-overlap-CS}
\end{align}
Substituting Eq.~(\ref{eq:benchmark-overlap-CS}) into Eq.~(\ref{eq:benchmark-overlap-derivative}), we obtain
\begin{align}
  \frac{\dd}{\dd t} \abs{\braket{d_m(t)|d_{m'}(t)}}^2 &\geq -2\abs{g(m-m')}\sqrt{V_{\rm B}}\, \abs{\braket{d_m(t)|d_{m'}(t)}}\times \sqrt{ 1-\abs{\braket{d_m(t)|d_{m'}(t)}}^2 }.
\end{align}
Therefore,
\begin{align}
  \frac{\dd}{\dd t} \sqrt{ 1-\abs{\braket{d_m(t)|d_{m'}(t)}}^2 } &\leq \abs{g(m-m')}\sqrt{V_{\rm B}}\, \abs{\braket{d_m(t)|d_{m'}(t)}} \\
  &\leq \abs{g(m-m')}\sqrt{V_{\rm B}}.
\end{align}
Integrating from $0$ to $t$, we have
\begin{align}
  \sqrt{ 1-\abs{\braket{d_m(t)|d_{m'}(t)}}^2 } \leq \abs{g(m-m')}\sqrt{V_{\rm B}}\,t.
\end{align}
Thus,
\begin{align}
  1-\abs{\braket{d_m(t)|d_{m'}(t)}}^2 \leq g^2V_{\rm B}(m-m')^2t^2.
\end{align}
Therefore, we have shown the inequality (\ref{eq:benchmark-overlap-bound}).

Substituting Eq.~(\ref{eq:benchmark-overlap-bound}) into Eq.~(\ref{eq:benchmark-trace-X}) and using $\sqrt{x^2+y^2}\leq \abs{x}+\abs{y}$, we obtain
\begin{align}
  \Tr X &\leq \frac12 \left[ f_1+f_K +\sum_{j=1}^{K-1}\abs{f_{j+1}-f_j} \right] +\abs{g}\sqrt{V_{\rm B}}\,t \sum_{j=1}^{K-1} \sqrt{f_jf_{j+1}} \left(\mu_{j+1}-\mu_j\right). \label{eq:benchmark-trace-X-overlap}
\end{align}
Since the sequence of $\{f_j\}:=\exp\qty[-\abs{\mu_j-\braket{s_0|O_{\rm S}|s_0}}/a]$ increases up to its maximum and then decreases as $j$ increases, the first term in Eq.~(\ref{eq:benchmark-trace-X-overlap}) can be bounded as follows:
\begin{align}
  \frac12 \left[ f_1+f_K +\sum_{j=1}^{K-1}\abs{f_{j+1}-f_j} \right] =\frac12\qty[f_1+f_K+2\max_j f_j-f_1-f_K] =\max_j f_j \leq 1. \label{eq:benchmark-weight-variation}
\end{align}

We next bound the second term in Eq.~(\ref{eq:benchmark-trace-X-overlap}).
Since the function $\exp\qty[-\abs{x-\braket{s_0|O_{\rm S}|s_0}}/a]$ is convex for an interval $[\mu_j,\mu_{j+1}]$ that does not contain $\braket{s_0|O_{\rm S}|s_0}$, we use the Hermite--Hadamard inequality and obtain the following bound:
\begin{align}
  \sqrt{f_jf_{j+1}} \left(\mu_{j+1}-\mu_j\right) &= \exp\left[ -\frac{ \abs{\mu_j-\braket{s_0|O_{\rm S}|s_0}} +\abs{\mu_{j+1}-\braket{s_0|O_{\rm S}|s_0}} }{2a} \right] \left(\mu_{j+1}-\mu_j\right) \\
  &\leq \int_{\mu_j}^{\mu_{j+1}} \exp\left( -\frac{\abs{x-\braket{s_0|O_{\rm S}|s_0}}}{a} \right)\dd x\\
  &\leq \int_{-\infty}^{\infty} \exp\left( -\frac{\abs{x-\braket{s_0|O_{\rm S}|s_0}}}{a} \right)\dd x =2a
\end{align}
There is at most one interval containing $\braket{s_0|O_{\rm S}|s_0}$ in its interior.
For such an interval, we have
\begin{align}
  \sqrt{f_jf_{j+1}} \left(\mu_{j+1}-\mu_j\right) &= \left(\mu_{j+1}-\mu_j\right) \exp\left[ -\frac{\mu_{j+1}-\mu_j}{2a} \right] \leq \frac{2a}{\me}.
\end{align}
Summing $\sqrt{f_jf_{j+1}}(\mu_{j+1}-\mu_j)$ over $j$, we obtain
\begin{align}
  \sum_{j=1}^{K-1} \sqrt{f_jf_{j+1}} \left(\mu_{j+1}-\mu_j\right) \leq \left(2+\frac{2}{\me}\right)a. \label{eq:benchmark-weighted-gap}
\end{align}
Combining Eqs.~(\ref{eq:benchmark-trace-X-overlap}), (\ref{eq:benchmark-weight-variation}), and (\ref{eq:benchmark-weighted-gap}), we finally obtain
\begin{align}
  \Tr X \leq 1+ \left(2+\frac{2}{\me}\right) \abs{g}\sqrt{V_{\rm B}}\,ta. \label{eq:benchmark-trace-X-bound}
\end{align}

We next bound the second term in Eq.~(\ref{eq:benchmark-entropy-X}).
From the definition of $f_j$, we have
\begin{align}
  -\sum_{j=1}^{K}p_{\mu_j}\log f_j &= \frac1a \sum_{j=1}^{K}p_{\mu_j}\abs{\mu_j-\braket{s_0|O_{\rm S}|s_0}} \leq \frac1a \sqrt{ \sum_{j=1}^{K}p_{\mu_j}(\mu_j-\braket{s_0|O_{\rm S}|s_0})^2 } = \frac{\Delta O_{\rm S}}{a}, \label{eq:benchmark-distribution-bound}
\end{align}
where we used the Cauchy--Schwarz inequality.
Substituting Eqs.~(\ref{eq:benchmark-trace-X-bound}) and (\ref{eq:benchmark-distribution-bound}) into Eq.~(\ref{eq:benchmark-entropy-X}), and using Eq.~(\ref{eq:benchmark-pure-qmi}), gives
\begin{align}
  I_{\rm S:D}(t) \leq 2\inf_{a>0} \left[ \log\left( 1+(2+2/\me)|g|\sqrt{V_{\rm B}}\,ta \right) +\frac{\Delta O_{\rm S}}{a} \right]. \label{eq:benchmark-regular-entropy-bound}
\end{align}

For $|g|\sqrt{V_{\rm B}}\,t\Delta O_{\rm S}>0$, choosing
\begin{align}
  a = \sqrt{ \frac{\Delta O_{\rm S}} {|g|\sqrt{V_{\rm B}}\,t} },
\end{align}
we have
\begin{align}
  I_{\rm S:D}(t) &\leq 2 \left[ \log\left( 1+ C\sqrt{ |g|\sqrt{V_{\rm B}}\, \Delta O_{\rm S}t } \right) \right. \left. + \sqrt{ |g|\sqrt{V_{\rm B}}\, \Delta O_{\rm S}t } \right] \\
  &\leq 2(C+1) \sqrt{ |g|\sqrt{V_{\rm B}}\, \Delta O_{\rm S}t }, \label{eq:benchmark-simple-qmi-bound}
\end{align}
where we used $\log(1+y)\leq y$ for $y\geq0$ and $C:=2+2/\me$.

For a fixed threshold $I_0>0$, let us define the correlation time: 
\begin{align}
  t_{\rm corr} := \inf\left\{ t>0: I_{\rm S:D}(t)\geq I_0 \right\}.
\end{align}
Whenever the threshold is reached, Eq.~(\ref{eq:benchmark-simple-qmi-bound}) requires
\begin{align}
  I_0 \leq 2(C+1) \sqrt{ |g|\sqrt{V_{\rm B}}\, \Delta O_{\rm S}t }.
\end{align}
Therefore, we obtain a lower bound on the correlation time as follows:
\begin{align}
  t_{\rm corr} \geq \frac{I_0^2} {4(C+1)^2 |g|\sqrt{V_{\rm B}}\, \Delta O_{\rm S}}. \label{eq:benchmark-correlation-time-bound}
\end{align}

\section{Derivation of Eq.~(\ref{main-eq:detector-coherent-state})}
We here derive the detector state given by Eq.~(\ref{main-eq:detector-coherent-state}).
Writing $H_0:=H_{\rm S}\otimes I_{\rm D}+I_{\rm S}\otimes H_{\rm D}$ and $H_{\rm int}:=O_{\rm S}\otimes \sum_k g_k(b^\dagger_k+b_k)$, and using the QND condition $[H_{\rm S},O_{\rm S}]=0$, we have the interaction Hamiltonian in the interaction picture as follows:
\begin{align}
  H^{\rm I}_{\rm int}(t) &=\me^{\mi H_0t}H_{\rm int}\me^{-\mi H_0t}\\
  &=\me^{\mi(I_{\rm S}\otimes H_{\rm D})t} H_{\rm int} \me^{-\mi(I_{\rm S}\otimes H_{\rm D})t}\\
  &=O_{\rm S}\otimes\sum_k g_k (b_k^\dagger\me^{\mi\omega_k t} +b_k\me^{-\mi\omega_k t}),
\end{align}
where we used $[H_{\rm S},O_{\rm S}]=0$, the Baker-Campbell-Hausdorff formula, and the bosonic commutation relations.
The time-evolution operator is
\begin{gather}
  U^{\rm I}(t)=\mathcal{T}\exp\qty[-\mi\int_0^t\dd s\,H_{\rm int}^{\rm I}(s)],
\end{gather}
where $\mathcal{T}$ is the time-ordering operator.
Using the Magnus expansion \cite{blanes2009magnus}, we write the time-ordered exponential as
\begin{gather}
  U^{\rm I}(t)=\exp[\sum_{n=1}^\infty \Omega_n(t)]\\
  \Omega_1(t)=-\mi\int_0^t\dd t_1\,H_{\rm int}^{\rm I}(t_1),\\
  \Omega_2(t)=-\frac{1}{2}\int_0^t\dd t_1 \int_0^{t_1}\dd t_2\, [H_{\rm int}^{\rm I}(t_1),H_{\rm int}^{\rm I}(t_2)],\\
  \Omega_3(t)=\text{Terms with triple commutators, } \cdots .
\end{gather}
The commutator between the interaction Hamiltonians at different times is given by
\begin{align}
  \qty[H_{\rm int}^{\rm I}(t_1),H_{\rm int}^{\rm I}(t_2)]&=O_{\rm S}^2\otimes \sum_{k,k'} g_k g_{k'} \Big[b_k^\dagger\me^{\mi\omega_k t_1} +b_k\me^{-\mi\omega_k t_1}, b_{k'}^\dagger\me^{\mi\omega_{k'} t_2} +b_{k'}\me^{-\mi\omega_{k'} t_2}\Big]\\
  &=O_{\rm S}^2\otimes \sum_k \abs{g_k}^2 \Big(\me^{\mi\omega_k(t_1-t_2)} \qty[b_k^\dagger,b_k] +\me^{-\mi\omega_k(t_1-t_2)} \qty[b_k,b_k^\dagger]\Big)\\
  &=O_{\rm S}^2\otimes \sum_k \abs{g_k}^2 \Big(\me^{-\mi\omega_k(t_1-t_2)} -\me^{\mi\omega_k(t_1-t_2)}\Big)I_{\rm D},
\end{align}
which is proportional to the identity in the detector Hilbert space and commutes with $H_{\rm int}^{\rm I}(t)$ for all $t$.
Therefore, all Magnus terms of order higher than the second vanish, and therefore the time-evolution operator exactly reduces to
\begin{gather}
  U^{\rm I}(t)=\exp\qty[\Omega_1(t)+\Omega_2(t)],\\
  \Omega_1(t)=-\mi O_{\rm S}\otimes\sum_k g_k \int_0^t \dd t_1 (b_k^\dagger\me^{\mi\omega_k t_1} +b_k\me^{-\mi\omega_k t_1}),\\
  \Omega_2(t)=-\frac{1}{2}O_{\rm S}^2\otimes\sum_k\abs{g_k}^2 \int_0^t \dd t_1 \int_0^{t_1} \dd t_2\, \times\qty(\me^{-\mi\omega_k(t_1-t_2)} -\me^{\mi\omega_k(t_1-t_2)})I_{\rm D}.
\end{gather}
Carrying out the integrals, we obtain
\begin{gather}
  \Omega_1(t)=O_{\rm S}\otimes\sum_k g_k \qty[\alpha_k(t)b_k^\dagger-\alpha_k^*(t)b_k],\\
  \Omega_2(t)=\mi\Psi(t)O_{\rm S}^2\otimes I_{\rm D},
\end{gather}
where
\begin{gather}
  \alpha_k(t):=\frac{1-\me^{\mi\omega_k t}}{\omega_k},\quad \Psi(t):=\sum_k\abs{g_k}^2 \frac{\omega_k t-\sin(\omega_k t)}{\omega_k^2}.
\end{gather}
Since $\Omega_1(t)$ and $\Omega_2(t)$ commute with each other, the evolution operator factorizes as
\begin{align}
  U^{\rm I}(t)&=\exp\qty[\Omega_1(t)]\exp\qty[\Omega_2(t)].
\end{align}
Note that $\exp[\Omega_1(t)]$ is a multi-mode displacement operator in the sense of quantum optics \cite{glauber1963coherent,sudarshan1963equivalence,mandel1995optical}.
Therefore, we obtain the time-evolution operator in the interaction picture as follows:
\begin{align}
  U^{\rm I}(t) &=\exp\!\left\{O_{\rm S}\otimes\sum_k g_k \left[\alpha_k(t)b_k^\dagger-\alpha_k^*(t)b_k\right]\right\}\notag\\
  &\quad\times\exp\!\left[\mi O_{\rm S}^2\otimes\Psi(t)I_{\rm D}\right], \label{eq:interaction-evolution-operator}\\
  \alpha_k(t)&:=\frac{1-\me^{\mi\omega_k t}}{\omega_k},\quad \Psi(t):=\sum_k\abs{g_k}^2 \frac{\omega_k t-\sin(\omega_k t)}{\omega_k^2}.
\end{align}
Applying Eq.~(\ref{eq:interaction-evolution-operator}) to $ \ket{\phi(0)} =\ket{s_0}\otimes\ket{\rm vac}_{\rm D}. $, we obtain the total state at time $t$ in the interaction picture:
\begin{align}
  \ket{\phi(t)}^{\rm I}
%   &=\sum_m\me^{-\mi m^2\Psi(t)}P_m\ket{s_0}\otimes
%   \bigotimes_k\ket{m g_k\alpha_k(t)}_{\rm D}\notag\\
  &:=\sum_{m: p_m>0} \sqrt{p_m}\,\me^{\mi m^2\Psi(t)} \ket{s_m}_{\mS}\otimes \ket{\beta_m(t)}_{\rm D}, \label{eq:joint-state}
\end{align}
where $p_m:=\braket{s_0|P_m|s_0}$ with $\sum_m p_m=1$, $\ket{s_m}:=P_m\ket{s_0}/\sqrt{p_m}$, and $\ket{\beta_m(t)}_{\rm D}$ is a multimode coherent state of the detector, whose mode-$k$ amplitude is $m g_k\alpha_k(t)$:
\begin{gather}
  \ket{\beta_m(t)}_{\rm D}:=\exp\qty[ m \sum_k g_k \qty(\alpha_k(t) b^\dagger_k - \alpha_k^*(t) b_k)]\ket{\rm vac}_{\rm D}.
\end{gather}
Tracing out the qubit ensemble and using the orthogonality of $\ket{s_m}$, we directly obtain the reduced state of the detector as follows:
\begin{align}
  \rho_{\rm D}^{\rm I}(t)&=\sum_m p_m \ketbra{\beta_m(t)}_{\rm D}.
\end{align}

\section{Evaluation of $A(t)$ with finite cutoffs}
Let us consider the case in which the detector has a finite frequency range, which can be modeled by an ultraviolet wave-number cutoff $\Lambda$.
With the cutoff, Eq.~(\ref{main-eq:main-A-cont}) becomes
\begin{align}
  A(t) &=\abs{g}^2S_D\int_{0}^{\Lambda} \dd k\, k^{D-1}\frac{2-2\cos(\Omega k^nt)}{\Omega^2k^{2n}}.
\end{align}
In the very short-time regime $t\ll (\Omega \Lambda^n)^{-1}$, $A(t)$ behaves as
\begin{gather}
  A(t)\simeq\abs{g}^2S_D\int_{0}^{\Lambda} \dd k\, k^{D-1}t^2 =\frac{\abs{g}^2S_D}{D}\Lambda^{D}t^2.
\end{gather}
In the intermediate-time regime $(\Omega \Lambda^n)^{-1}\ll t$, the change of variable $u=\Omega k^nt$ gives
\begin{align}
  A(t)&=\frac{2\abs{g}^2S_D}{n\Omega^\nu}t^{2-\nu} \int_{0}^{\Omega\Lambda^nt} \dd u\,u^{\nu-3}(1-\cos u), \label{eq:end-A-u}
\end{align}
where $\nu=D/n$, and $\Lambda$ is the ultraviolet wave-number cutoff.
For $0<\nu<2$, it follows that
\begin{align}
  A(t)&=\frac{2\abs{g}^2S_D}{n\Omega^\nu}t^{2-\nu} \Bigg\{\frac{\pi}{2\Gamma(3-\nu)\sin(\pi \nu/2)} +\mathcal{O}[(\Omega\Lambda^nt)^{\nu-2}]\Bigg\}.
\end{align}
Therefore, if the finite cutoff $\Lambda$ exists, the fractional behavior $A(t)\propto t^{2-\nu}$ is still valid for times $t\gg (\Omega \Lambda^n)^{-1}$.

\section{Photon-number measurement}
We here show the error of the number operator \eqref{main-eq:main-error}.
Let us calculate the expectation value of the number operator $\mathcal{N}(t):=B_0^\dagger(t) B_0(t)$.
Since the detector coherent states (\ref{main-eq:protocol-env}) are eigenstates of the annihilation operator $B_0(t)$ with eigenvalue $(2\ell-N)\sqrt{A(t)}$, we have
\begin{align}
  \braket{\mathcal{N}(t)} &:=\Tr_{\rm D}\qty[\rho^{\rm I}_{\rm D}(t)\mathcal{N}(t)] \\
  &=A(t)\sum_{\ell=0}^N q_\ell(2\ell-N)^2\\
  &=A(t)\sum_{\ell=0}^N q_\ell \qty(4\ell^2-4N\ell+N^2),\label{eq:end-number-mean}
\end{align}
where $q_\ell$ is defined in Eq.~(\ref{main-eq:protocol-env}).
Setting $p=\sin^2\theta$, we have the moments of this binomial distribution \cite{johnson2005univariate} as follows:
\begin{align}
  \sum_{\ell=0}^Nq_\ell \ell&=Np,\label{eq:end-binomial-first-one}\\
  \sum_{\ell=0}^Nq_\ell \ell^2&=N(N-1)p^2+Np. \label{eq:end-binomial-first-two}
\end{align}
Using Eqs.~(\ref{eq:end-binomial-first-one}) and (\ref{eq:end-binomial-first-two}) in Eq.~(\ref{eq:end-number-mean}), we obtain
\begin{align}
  \braket{\mathcal{N}(t)} &=A(t)\qty{4\qty[N(N-1)p^2+Np] -4N(Np)+N^2}\\
  &=A(t)\qty[N^2-N(N-1)\sin^2(2\theta)].\label{eq:end-number-mean-final}
\end{align}

Next, we calculate the second moment of the number operator:
\begin{align}
  \mathcal{N}^2(t) &=\qty[B_0^\dagger(t)B_0(t)]^2=\qty[B_0^\dagger(t)]^2B_0^2(t)+\mathcal{N}(t).
\end{align}
Then the second moment is given by
\begin{align}
  \braket{\mathcal{N}^2(t)} &:=\Tr_{\rm D}\qty[\rho^{\rm I}_{\rm D}(t)\mathcal{N}^2(t)]\\
  &=A^2(t)\sum_{\ell=0}^Nq_\ell(2\ell-N)^4 +A(t)\sum_{\ell=0}^Nq_\ell(2\ell-N)^2\\
  &=A^2(t)\Bigg[16\sum_{\ell=0}^Nq_\ell \ell^4 -32N\sum_{\ell=0}^Nq_\ell \ell^3+24N^2\sum_{\ell=0}^Nq_\ell \ell^2-8N^3\sum_{\ell=0}^Nq_\ell \ell+N^4\Bigg] +\braket{\mathcal{N}(t)}. \label{eq:end-number-second}
\end{align}
The second term on the right-hand side of Eq.~(\ref{eq:end-number-second}) is equal to $\braket{\mathcal{N}(t)}$.
To evaluate the first term, we use the properties of the higher moments of the binomial distribution \cite{johnson2005univariate} as follows:
\begin{align}
  \sum_{\ell=0}^Nq_\ell \ell^3 &=N(N-1)(N-2)p^3 +3N(N-1)p^2+Np, \label{eq:end-binomial-third}\\
  \sum_{\ell=0}^Nq_\ell \ell^4 &=N(N-1)(N-2)(N-3)p^4 +6N(N-1)(N-2)p^3 +7N(N-1)p^2+Np. \label{eq:end-binomial-higher}
\end{align}
Substituting Eqs.~(\ref{eq:end-binomial-first-one}), (\ref{eq:end-binomial-first-two}), (\ref{eq:end-binomial-third}), and (\ref{eq:end-binomial-higher}) into Eq.~(\ref{eq:end-number-second}) and using Eq.~(\ref{eq:end-number-mean-final}), we obtain
\begin{align}
  \braket{\mathcal{N}^2(t)} &=A^2(t)\Bigg\{16\Big[N(N-1)(N-2)(N-3)p^4 +6N(N-1)(N-2)p^3 +7N(N-1)p^2+Np\Big]\notag\\
  &\quad-32N\Big[N(N-1)(N-2)p^3 +3N(N-1)p^2+Np\Big]+24N^2\Big[N(N-1)p^2+Np\Big] -8N^4p+N^4\Bigg\}\notag\\
  &\quad+A(t)\qty[N^2-N(N-1)\sin^2(2\theta)]. \label{eq:end-number-second-expanded}
\end{align}
Therefore, combining Eqs.~(\ref{eq:end-number-mean-final}) and (\ref{eq:end-number-second-expanded}), the variance of the number operator is given by
\begin{align}
  \Delta^2\mathcal{N}(t) &:=\braket{\mathcal{N}^2(t)} -\braket{\mathcal{N}(t)}^2\\
  &=A^2(t)N(N-1)\sin^2(2\theta) \times\qty[(2N-1)+(2N-3)\cos(4\theta)]+A(t)\qty[N^2-N(N-1)\sin^2(2\theta)].
\end{align}

We therefore consider the error for estimating $\theta$ as follows:
\begin{align}
  \Delta^2\theta&:=\Delta^2\mathcal{N}(t)/\abs{\partial_{\theta}\braket{\mathcal{N}(t)}}^2\\
  &=\frac{[(2N-1)+(2N-3)\cos(4\theta)] \sin^2(2\theta)}{4N(N-1)\sin^2(4\theta)}+\frac{N^2-N(N-1)\sin^2(2\theta)} {4A(t)N^2(N-1)^2\sin^2(4\theta)}. \label{eq:main-error-exact}
\end{align}
Note that we assume that $\sin(4\theta)\neq 0$ to avoid the divergence.
For large $N$, its sensitivity is approximated by
\begin{align}
  \Delta^2\theta &\simeq\frac{1} {4N}+ \frac{1}{16A(t)N^2\sin^2(2\theta)}. \label{eq:main-error}
\end{align}

\bibliography{rapidreadout}